\pdfoutput=1 
\documentclass[conference]{IEEEtran}
\IEEEoverridecommandlockouts
\usepackage[bottom]{footmisc}% places footnotes at page bottom
\usepackage{amsmath,amssymb} 
\usepackage{float}
\usepackage{multirow}
\usepackage{graphicx}
\usepackage{color}
\usepackage{booktabs}
\usepackage{todonotes}
\usepackage[hidelinks]{hyperref}
\usepackage[capitalise]{cleveref}
\usepackage{cite} 
\usepackage{threeparttable}
\usepackage{makecell}
\usepackage{listings}
\usepackage{glossaries}
\usepackage{siunitx}
\usepackage{enumitem}
\usepackage[table, dvipsnames]{xcolor}
\usepackage{colortbl}
\usepackage{amsfonts}
\usepackage{tikz}
\usepackage{subcaption} 
\usepackage{flushend}
\usepackage{array}
\usepackage{pgfplots}
\usepackage{pgfplotstable}
\pgfplotsset{compat=1.18}
\pgfplotsset{compat=newest}
\usepackage{balance}

\usepackage{subcaption} % in preamble

\usepackage{amsmath}
\usepackage{algorithm}
\usepackage{algpseudocode}

\usepackage{tikz}
\usetikzlibrary{arrows.meta, positioning}
\usetikzlibrary{calc} % <— needed for coordinate math
\usepackage{sansmath} % optional: nicer sans ticks
\usepgfplotslibrary{groupplots}
\usepackage{pgfplots}
\pgfplotsset{compat=1.18}

\definecolor{codegreen}{rgb}{0,0.6,0}
\definecolor{codegray}{rgb}{0.5,0.5,0.5}
\definecolor{codepurple}{rgb}{0.58,0,0.82}
\definecolor{backcolour}{rgb}{0.95,0.95,0.92}

\definecolor{colordt}{rgb}{1.0, 0.0, 0,22}  % carnelian 
\definecolor{colormlp}{rgb}{0.94, 0.86, 0.51}  % buff
\definecolor{colorsvm}{rgb}{0.0, 0.28, 0.67}  % cobalt
\definecolor{colorbnn}{rgb}{0.29, 0.33, 0.13} % army green
\definecolor{colortnn}{rgb}{0.43, 0.21, 0.1} % auburn

\definecolor{areaADC}{RGB}{51,102,255} % Blue
\definecolor{areaFEx}{RGB}{102,153,255} % Lighter Blue
\definecolor{areaClf}{RGB}{153,204,255} % Lightest Blue
\definecolor{powerADC}{RGB}{255,51,51} % Red
\definecolor{powerFEx}{RGB}{255,102,102} % Lighter Red
\definecolor{powerClf}{RGB}{255,153,153} % Lightest Red

\definecolor{lightgreen}{RGB}{220,255,220}

\lstdefinestyle{mystyle}{
    backgroundcolor=\color{backcolour},   % Set background color
    commentstyle=\color{codegreen},        % Style for comments
    keywordstyle=\color{blue},             % Style for keywords
    numberstyle=\tiny\color{codegray},     % Style for line numbers
    stringstyle=\color{codepurple},        % Style for strings
    basicstyle=\ttfamily\small,            % Basic font style
    breakatwhitespace=false,               % Do not break lines at whitespace
    breaklines=true,                       % Enable line breaking
    captionpos=b,                          % Caption position at bottom
    keepspaces=true,                       % Keep spaces in text
    numbers=left,                          % Line numbers on the left
    numbersep=5pt,                         % Distance of line numbers from code
    showspaces=false,                      % Do not show spaces
    showstringspaces=false,                % Do not show spaces in strings
    showtabs=false,                        % Do not show tabs
    tabsize=2,                             % Set tab size to 2 spaces
    language=Python                        % Set language to Python
}

\newcommand{\red}[1]{{\color{black}#1}}

\colorlet{mc}{LimeGreen!50!White!50!}

\usepackage{soul}

\def\BibTeX{{\rm B\kern-.05em{\sc i\kern-.025em b}\kern-.08em
    T\kern-.1667em\lower.7ex\hbox{E}\kern-.125emX}}

\begin{document}
\bstctlcite{IEEEexample:BSTcontrol}

\title{

{\vspace{-0.7cm}\small This article is accepted for publication in \textit{IEEE Design, Automation \& Test in Europe in Europe (DATE), 2026 } \\
% This is the author’s version which has not been fully edited, and content may change prior to final publication. \\
}
\vspace{-0.9\baselineskip}
\rule{\textwidth}{0.4pt}\\
Design and Optimization of Mixed-Kernel Mixed-Signal SVMs for Flexible Electronics
% \vspace{-1ex}
}
% \title{Design and Optimization of Mixed-Kernel Mixed-Signal SVMs for Flexible Electronics}

\author{
    \IEEEauthorblockN{
        Florentia Afentaki\IEEEauthorrefmark{1}$^{\circ}$,
        Maha Shatta\IEEEauthorrefmark{2}$^{\circ}$,
        Konstantinos Balaskas\IEEEauthorrefmark{1},
         Georgios Panagopoulos\IEEEauthorrefmark{9},\\
        Georgios Zervakis\IEEEauthorrefmark{1},
        Mehdi B. Tahoori\IEEEauthorrefmark{2}
     \thanks{$^{\circ}$Authors contributed equally to this work.}
    }
    
    \IEEEauthorblockA{\IEEEauthorrefmark{1}University of Patras, GR
    \IEEEauthorrefmark{2}Karlsruhe Institute of Technology, DE
      \IEEEauthorrefmark{9}National Technical University of Athens, GR
    }

    \IEEEauthorblockA{
    \IEEEauthorrefmark{1}\{afentaki, kompalas, zervakis\}@ceid.upatras.gr
    \IEEEauthorrefmark{2}\{maha.shatta, mehdi.tahoori\}@kit.edu
    \IEEEauthorrefmark{9}gepanago@mail.ntua.gr
    }
    
}
\vspace{-6ex}
 
\maketitle

\begin{abstract}
Flexible Electronics (FE) have emerged as a promising alternative to silicon-based technologies, offering on-demand low-cost fabrication, conformality, and sustainability. 
However, their large feature sizes severely limit integration density, imposing strict area and power constraints, thus prohibiting the realization of Machine Learning (ML) circuits, which can significantly enhance the capabilities of relevant near-sensor applications.
Support Vector Machines (SVMs) offer high accuracy in such applications at relatively low computational complexity, satisfying FE technologies' constraints.
Existing SVM designs rely solely on linear or Radial Basis Function (RBF) kernels, forcing a trade-off between hardware costs and accuracy.
Linear kernels, implemented digitally, minimize overhead but sacrifice performance, while the more accurate RBF kernels are prohibitively large in digital, and their analog realization contains inherent functional approximation.
% , while RBF kernels offer higher accuracy, but are prohibitively large in digital, while their more efficient offer higher accuracy at reasonable cost, but inherently contain functional approximations.
In this work, we propose the first mixed-kernel and mixed-signal SVM design in FE,
% balancing the cost vs accuracy trade-off 
which unifies the advantages of both implementations and balances the cost/accuracy trade-off.
% combining the hardware efficiency of digital linear kernels with highly-accurate analog RBF ones.
% Furthermore, we present a co-optimized training and mapping methodology that jointly determines kernel assignments and hardware deployment.
To that end, we introduce a co-optimization approach that trains our mixed-kernel SVMs and maps binary SVM classifiers to the appropriate kernel (linear/RBF) and domain (digital/analog), aiming to maximize accuracy whilst reducing the number of costly RBF classifiers.
% Thus, balancing accuracy with hardware efficiency whilst meeting the stringent constraints of FE.
Our designs deliver \red{7.7\%} higher accuracy than state-of-the-art single-kernel linear SVMs, and reduce area and power by \red{$108\times$} and \red{$17\times$} on average compared to digital RBF implementations.
\end{abstract}


\section{Introduction}
\label{sec:introduction}

% \red{The growing accumulation of waste poses significant environmental and health risks, demanding attention from developers of emerging technologies~\cite{scagliarini2022soft}. 
% Beyond enabling new functionalities, modern systems are expected to align with stricter regulations on waste reduction and raw material consumption, while simultaneously responding to consumer demand for sustainable, “green” solutions.
% In this context, Flexible Electronics (FE) offer a promising pathway: they combining low-cost electronics with recyclable and environmentally benign materials, thereby supporting the drive toward sustainable electronics. }
Flexible Electronics (FE) have emerged as a promising alternative to silicon-based computing, offering attractive properties such as mechanical flexibility, non-toxicity, and conformality, whilst enabling low-cost fabrication and sustainability. 
These characteristics make FE suitable for low-cost far-edge applications, such as wearable healthcare monitoring~\cite{xu:nature2014:epidermal, bodytemperature:sna:2020,pressuresensor:research:2022,wearable:adma:2022,Wearable:acssensors:2019,healthcare:Nanoscale:2024}, and smart packaging~\cite{smartpackaging2022, disposable:JSNB:2023}.
However, FE is fundamentally constrained by large feature sizes, limited device counts, and low integration density,
% Additionally, the absence of p-type devices enforces resistive-nMOS logic.
which result in increased power and area consumption, limiting scalability and efficiency.
Thus, implementing complex circuits in FE, such as machine learning (ML) classifiers, remains challenging.
Yet, ML-based classification is essential for sensor-driven applications that translate raw signals into meaningful decisions, which are central to FE use cases~\cite{Mubarik:MICRO:2020:printedml}. 
% , even though ML-based classification is central to relevant sensor-driven applications~\cite{Mubarik:MICRO:2020:printedml}.

% Thus, the realization of complex circuits in FE, such as machine learning (ML) algorithms, is particularly challenging~\cite{Mubarik:MICRO:2020:printedml}.
% Integrating Machine Learning (ML) algorithms into such applications is particularly appealing~\cite{Mubarik:MICRO:2020:printedml}.
% Unfortunately, implementing complex ML classifiers in FE remains challenging due to large feature sizes and limited device integration, which introduce considerable power and area overheads~\cite{ozer:nature2024:bendableRiscV}. 
% These constraints hinder the widespread deployment of ML in flexible systems.  

To address these limitations, most state-of-the-art FE designs rely on approximate computing---exploiting the inherent error tolerance of ML, trading accuracy for efficiency~\cite{Henkel:ICCAD2022:expedition,Armeniakos:DATE2022:axml,Armeniakos:TCAD2023:cross,Armeniakos:TC2023:codesign,Afentaki:ICCAD23:hollistic,Afentaki:DATE2024:embedding,Mrazek:ICCAD2024}. 
Typically, multi-layer perceptrons (MLPs) are employed, which require aggressive approximation to mitigate their large area/power overhead, and thus introduce large accuracy loss.
% and therefore depend on aggressive approximation, which degrades task-level accuracy.

Support vector machines (SVMs) are a favorable alternative to MLPs for the simpler classification tasks targeted by FE, offering modest model complexity, 
reduced area/power requirements, and competitive accuracy across a range of relevant applications~\cite{Burges:DMKD:1998,Smola:StatsComp:2004}.
At their core, SVMs typically rely on either linear or radial basis function (RBF) kernels.
RBF kernels provide higher classification accuracy~\cite{Cavallaro:TCAS:2010,Li:ISCAS:2017}---in our experiments, 7.7\% higher accuracy on average than linear kernels---but incur prohibitively large area and power overheads for realization in the digital domain.
Thus, RBF-based SVMs in FE can only include cost-effective analog circuits, which, however, contain inherent functional approximation.
On the other hand, linear kernels offer reduced computational complexity and hardware cost, but primarily result in lower achievable accuracy. 
Thus, restricting flexible SVMs to a single kernel type inevitably forces a compromise between classification accuracy and hardware efficiency.

In this work, we present the first mixed-kernel, mixed-signal SVM tailored for FE.
Specifically, we combine linear and RBF kernels in a unified design, aiming to achieve high accuracy while instantiating only a small number of RBF units.
% For multiclass tasks, the problem decomposes into binary decision functions (OvR/OvO). 
Leveraging that multiclass SVMs can be decomposed into binary classifiers, we implement the RBF classifiers as analog FlexICs---directly computing on sensory inputs---whereas linear ones as digital designs.
% , yielding a mixed-signal SVM that reduces hardware cost.
Operating in the subthreshold regime, analog RBF blocks are based on functionally approximate Gaussian kernels, enabling ultra-low-power and high area efficiency. 
Thus, our mixed-kernel design improves accuracy by selectively deploying RBF beyond linear-only baselines.
It also shifts the otherwise prohibitive digital RBF into the analog domain, reducing both power and area. 
Furthermore, we propose a separation-based exploration strategy that enables the automatic identification of the optimal kernel type per binary classifier---by assessing its accuracy contribution---and facilitates mixed-kernel SVM training optimized for high accuracy while minimizing the number of costly RBF classifiers.
Our designs achieve $7.7\%$ higher accuracy on average compared to the state-of-the-art single-kernel linear SVMs, with $108\times$ and $17\times$ average lower area and power, respectively, compared to all-RBF digital solutions.

\noindent\textbf{The main contributions of this work are as follows:}
\begin{enumerate}[topsep=0pt,leftmargin=*]
\item We propose the first mixed-signal mixed-kernel SVM design in FE, implementing RBF kernels in the analog domain and linear kernels in digital.%, enabling highly-accurate flexible SVMs.
\item We introduce a separation-driven exploration that trains mixed-kernel SVMs and systematically allocates classifiers to kernel types, achieving high accuracy.
% \item Our designs achieve \red{$2\%$} higher accuracy on average compared to single-kernel state-of-the-art SVMs, and \red{$32.5\times$} lower power on average compared to single-kernel RBF digital SVMs.
% whilst satisfying the tight constraints of FE.
% \item Our SVM designs achieve \textcolor{red}{$XX\%$} higher accuracy, with an area gain of \textcolor{red}{$XX\times$} and energy reduction of \textcolor{red}{$XX\times$}, on average, compared to the state of the art.
\end{enumerate}

\section{Background}
\label{sec:background}

\subsection{Support Vector Machine}
\label{sec:background:svms_alg}
SVMs are supervised classifiers---robust to overfitting and well-suited for small datasets---that identify a set of $m$ support vectors to determine an optimal hyperplane in feature space, using the following binary decision function for separating between two classes:
% Various kernels can be used to enhance classification (e.g., RBF, linear), each contributing distinct attributes and allowing SVMs to adapt to various data types and classification problems. 
\begin{equation}
    f(\mathbf{x})=\mathrm{sign}\!\left(\sum_{i=1}^{m}\alpha_i y_i\, K(\mathbf{x}_i,\mathbf{x}) + b\right),
\label{eq:svm_decision}
\end{equation}
where $\{(\mathbf{x}_i,y_i)\}_{i=1}^m$ are support vector data with $y_i\in\{-1,+1\}$, $\alpha_i\!\ge 0$ are the dual coefficients, $b$ is the bias, and $K(\cdot,\cdot)$ is a kernel function \cite{Kang2010gaussian}.
For an RBF kernel:
\begin{equation}
K(\mathbf{w}_i,\mathbf{x})=\exp\!\big(-\gamma\|\mathbf{x}_i-\mathbf{x}\|^2\big), \quad \gamma>0,
\label{eq:rbf}
\end{equation}
whereas for linear: $K(\mathbf{x}_i,\mathbf{x})=\mathbf{x}_i^\top\mathbf{x}$, where $\mathbf{x}\in\mathbb{R}^D$ denotes the input features.
For the linear case, the dual and primal formulations coincide \cite{burges1998tutorial}, yielding a single weight vector:
\begin{equation}
\mathbf{w}=\sum_{i=1}^{m}\alpha_i y_i\,\mathbf{x}_i,
\qquad
f(\mathbf{x})=\mathrm{sign}\!\big(\mathbf{w}^\top\mathbf{x}+b\big).
\label{eq:linear_primal}
\end{equation}
% Hence, inference reduces to one dot product and a bias add, requiring $O(d)$ MACs, no exponentials, and no dependence on support vectors $N$ at run time. 
% This also simplifies storage to $(\mathbf{w},b)$ only. 
% With the RBF kernel, inference for a new $\mathbf{x}\in\mathbb{R}^d$ requires, for each support vector, computing a squared Euclidean distance (an $O(d)$ multiply–accumulate, MAC, pattern), followed by an exponential and a weighted accumulation via $\alpha_i y_i$. 
% The overall complexity is $O(N_{\mathrm{SV}}\,d)$ MACs plus $N_{\mathrm{SV}}$ nonlinear evaluations, where $N_{\mathrm{SV}}$ is the number of support vectors retained by training; memory scales with the storage of all support vectors and their dual coefficients.
Linear kernels reduce inference to one dot product and bias addition, requiring $O(D)$ multiply-accumulate (MAC) operations with only one weight parameter per input and no dependence on support vectors $m$ at run time, compared to $O(m \cdot D)$ for RBF.
% For the linear kernel, the dual and primal formulations coincide, yielding a single weight vector
% \begin{equation}
% \mathbf{w}=\sum_{i=1}^{N}\alpha_i y_i\,\mathbf{x}_i,
% \qquad
% f(\mathbf{x})=\mathrm{sign}\!\big(\mathbf{w}^\top\mathbf{x}+b\big).
% \label{eq:linear_primal}
% \end{equation}
% Hence, inference reduces to one dot product and a bias add, requiring $O(d)$ MACs, no exponentials, and no dependence on $N_{\mathrm{SV}}$ at run time. 
% This also simplifies storage to $(\mathbf{w},b)$ only. 
In practice, RBF kernels can offer stronger non-linear decision boundaries at the cost of higher computation (for euclidean distance and exponential calculation), whereas linear SVMs trade robustness for a simplified, low-latency datapath.
We extend SVMs to $K$-class problems via the One-vs-One (OvO) strategy. 
% OvO typically attains high accuracy for targeted applications~\cite{sertaridis:ISCAS2025:SequentialSVM_1_mac}.
In OvO, each binary classifier discriminates between a pair of classes, outputting either 0 or 1 indicating the winner, with predictions made by majority voting over all outputs.
As explained later, this property (i.e., analog-in, digital-out) considerably simplifies the integration of analog kernels in our mixed-signal architecture.
% Thus, OvO is used in our work hereafter.

% fixed, hardware-friendly datapath consisting solely of a $d$-term MAC and an add, with deterministic latency and energy that scale only with the input dimension $d$.

% SVMs can be extended to $K$-class problems via the One-vs-Rest (OvR) and One-vs-One (OvO) strategies. 
% Even though OvR trains only $K$ binary classifiers to OvO's $\binom{K}{2}$, and is typically more area-efficient, OvO typically attains higher accuracy than OvR for targeted applications~\cite{sertaridis:ISCAS2025:SequentialSVM_1_mac}.
% each separating one class from the rest, and predicts by selecting the classifier with the largest (signed) margin. 
% Each OvO binary classifier discriminates between a pair of classes, outputting either 0 or 1 indicating the winner, with predictions made by majority voting over all outputs.
% As explained, later, this property considerably simplifies the integration of analog kernels in our mixed-signal architecture.
% Thus, OvO is used in our work hereafter.

\subsection{Flexible Electronics}
\label{sec:background:flexible_electronics}
% \todo[inline]{Figure for technology}
Recent advancements in FE involve the design of Flexible Integrated Circuits (FlexICs) with Indium Gallium Zinc Oxide (IGZO) Thin-Film Transistors (TFTs), offering mechanical adaptability and flexibility with cost-effective manufacturing~\cite{ozer:nature2024:bendableRiscV}.
Unlike conventional silicon devices, IGZO TFTs can be made on lightweight, flexible substrates (e.g., polyimide) using low-temperature lithography and without needing protective packaging.
% This approach avoids rigid silicon wafers and high-temperature processing, 
Thus, they substantially lower production costs, fabrication time (from 32 weeks to under 3.5~\cite{pragmatic:whitepaper:sustainability}), and environmental impact (e.g., water consumption, carbon emissions). 
% IGZO TFTs also offer inherent mechanical flexibility, enabling bending without extra encapsulation. 
% Their streamlined process also reduces fabrication time from 32 weeks to under 3.5 days, supporting scalable deployment~\cite{pragmatic:whitepaper:sustainability}.
Despite these benefits, IGZO TFTs fall behind CMOS in performance and feature size (e.g., \SI{600}{\nano\meter} for FlexICs compared to a few \si{\nano\meter} in silicon~\cite{ozer:nature2024:bendableRiscV}).
Moreover, the technology provides only n-type devices, restricting designs to unipolar logic and necessitating resistor-NMOS (R-NMOS) circuits, where a pull-up resistor replaces the pMOS transistor.
This directly impacts delay and power consumption, and poses significant challenges for designing complex circuits, such as ML classifiers.
% Consequently, implementing complex circuits—such as ML classifiers—for area-constrained applications like wearables is challenging in FE. 
Our bespoke mixed-signal flexible SVM circuits combat these inherent challenges, by offloading computationally-demanding RBF kernels in the analog domain, reducing both area and power significantly compared to digital deployment.
% and eliminating the need for costly memory elements with fully-unrolled designs.
% To address these issues, hardware–software co-design optimizations are essential in FE~\cite{Ozer:2019:Bespoke}. 
% Our approach optimize the flexible SVM classifier, reducing memory usage, gate count, and power consumption, while also eliminating the need for memory elements that are scarce and costly in FE technology—key enablers for efficient, lightweight flexible classification systems.
% \todo[inline]{bespoke, architectural difference (mixed-signal), high-level kernel selection}

\subsection{Related Work}
\label{sec:background:related}
Several SVM hardware implementations have been proposed, mostly relying on CMOS technology.
% have been proposed, targeting digital, analog, and mixed-signal domains.
On the digital side,~\cite{afifi2020fpga} provides a review of SVM classifiers implemented on FPGA platforms. 
A practical example is given in~\cite{afifi:IECBES2016:fgpa_SVM_melanoma}, which demonstrates a low-cost FPGA-based SVM classifier for melanoma detection on a Xilinx Zynq device. 
In contrast, analog CMOS realizations target ultra-low-power operation, such as the fully trainable Gaussian-kernel SVM presented in~\cite{tcsi2009analogsvm}. 
More recently,~\cite{alimisis2023hybridsvm} reported a low-power analog integrated implementation of SVM with on-chip learning, tested on a bearing fault detection application. 
However, these prior works rely on rigid silicon-based technologies, which lack mechanical properties such as flexibility and conformability, and also do not face the inherent limitations of FE.
% Although not directly focused on SVMs, progress in flexible digital design flows has been demonstrated with complete microprocessors such as Flex6502~\cite{celiker2022flex6502}, highlighting the broader feasibility of complex circuits in emerging technologies.

In the context of printed and flexible electronics~\cite{tahoori:ETS2025:PFE}, SVM classifiers have been investigated primarily along two directions: (i) fully-parallel approximated architectures that reduce hardware cost at the expense of accuracy~\cite{Armeniakos:DATE2022:axml,Armeniakos:TCAD2023:cross}, and (ii) sequential architectures that significantly lower area but incur high energy consumption~\cite{sertaridis:ISCAS2025:SequentialSVM_1_mac,besias:DATE2025:Sequential_N_mac}.
In~\cite{Armeniakos:DATE2022:axml}, hardware-friendly weight replacement and gate-level pruning were proposed as post-training approximations, while~\cite{Armeniakos:TCAD2023:cross} extended this with voltage over-scaling at the circuit level. 
More recently,~\cite{sertaridis:ISCAS2025:SequentialSVM_1_mac,besias:DATE2025:Sequential_N_mac} further explored sequential SVM architectures, emphasizing the trade-off between area and energy consumption.
However, these implementations only consider linear kernels, avoiding the hardware cost and design complexity of digital RBF kernels but resulting in diminished accuracy---even more so in approaches that exploit approximate computing.

\section{Proposed Mixed-signal Mixed-kernel SVMs}
\label{sec:framework}

\subsection{Architecture Overview}
\label{sec:architecture}
\begin{figure}[!t]
\centering
\includegraphics[width=\columnwidth]{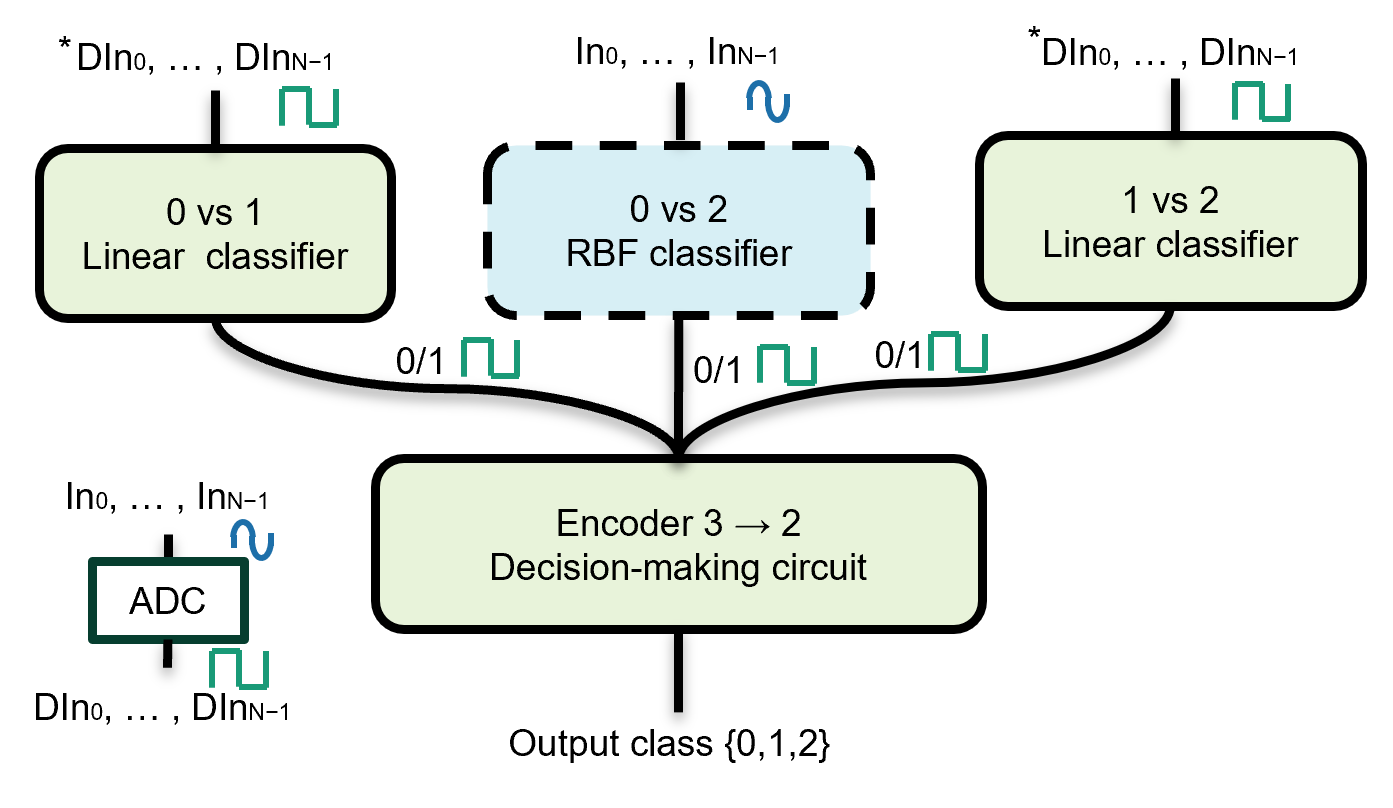}
\caption{Overview of the proposed mixed-kernel, mixed-signal SVM for three classes: RBF classifiers are realized in analog, while linear classifiers and decision logic are in digital.}
\label{fig:mixed_kernel_SVM}
\end{figure}

% \todo[inline]{ADC mention}
Fig.~\ref{fig:mixed_kernel_SVM} illustrates an example of our proposed mixed-signal mixed-kernel SVM architecture for three classes, which partitions linear and RBF kernel across digital and analog domains, respectively.
The linear classifiers require analog-to-digital converters (ADCs) to digitize their inputs, whereas the analog RBF classifiers operate directly on analog signals.
Per the OvO scheme, one binary classifier is instantiated for each two classes, yielding a single binary value (0 or 1, according to the winning class) for both RBF and linear classifiers.
Thus, as analog classifiers produce digital outputs, no additional conversion is needed, eliminating potential ADC costs.
This also 
% improves robustness to analog-induced noise, and 
considerably simplifies the result aggregation by the digital decision-making circuitry, responsible for encoding the predicted class.
All classifiers are realized in parallel, eliminating the need for control circuitry or sequential elements---costly in FE.
% Preamble:
% \usepackage{tikz}
% \usetikzlibrary{arrows.meta,positioning}

% \todo[inline]{highlight which inputs are digital and which not in the figure}
% \todo[inline]{decision making circuit }

% Nonlinear RBF kernels are assigned to the analog domain, where exponentials can be computed efficiently, while linear kernels together with the multiclass decision logic are realized in the digital domain.
% This separation enables accurate and efficient classification without incurring the overhead of an entirely digital design or the error sensitivity of a fully analog implementation.
% Integrating the decision-making circuit logic in the digital domain ensures robustness against noise and process variations.
% The rest of this chapter is organized as follows: we first describe the RBF classifier implementation in the analog domain, then present the linear classifiers and the decision-making logic in the digital domain, and finally discuss the high-level exploration that determines the optimal configuration of kernels—assigning them to either RBF (analog) or linear (digital)—with the goal of maximizing accuracy.

\subsection{Analog RBF Classifier}
\label{subsec:analog}

\begin{figure}[t]
\centering
  \includegraphics[width=\columnwidth]{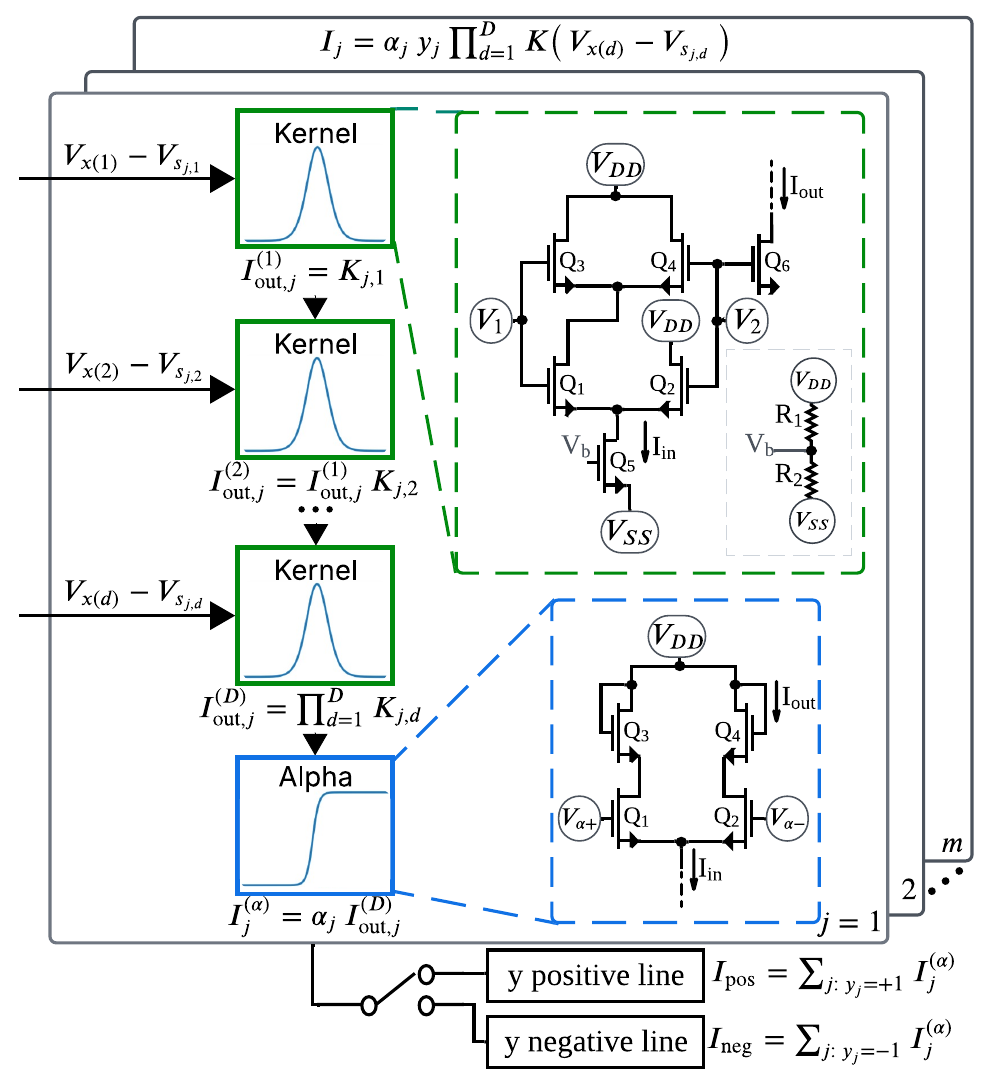} % placeholder
\caption{Analog RBF classifier architecture: a hardware Gaussian kernel forms a separable product across dimensions; outputs are weighted by \(\alpha\) and routed by the sign \(y\), then differentially accumulated across support vectors.}
\label{fig:analog_rbf_kernel}
\end{figure}

First, we describe the design of our analog RBF classifier.
Designing in FlexIC presents challenges compared to CMOS, as the technology includes only n-type transistors and lacks p-type devices.
Still, based on the CMOS kernel design of~\cite{Kang2010gaussian}, we implement the RBF classifier in our flexible technology as shown in Fig.~\ref{fig:analog_rbf_kernel}, using an n-type transistor differential pair biased in subthreshold, leveraging the exponential $I$–$V$ characteristic of this regime.
Subthreshold biasing enables ultra-low power, aligning with our target applications.
Even though the reduced bias currents may introduce longer settling time, this latency is effectively hidden in the mixed-signal design by the ADC conversion time in the digital path, providing timing slack for the analog stage.
Also, FE circuits typically operate within the Hz range, since performance is a lower priority to area and power efficiency.
% We adapted the design from~\cite{Kang2010gaussian}, showing the whole classifier.

\subsubsection{RBF kernel (Gaussian)}
The kernel is implemented with two subthreshold n-type transistors differential pairs, $(Q_1,Q_2)$ cascaded by $(Q_3,Q_4)$, and $Q_5$ for biasing the kernel and setting the operating point. The circuit produces a current-mode bell-shaped response that, near the origin, is well modeled by a squared hyperbolic secant.
Specifically, cascading the complementary pairs gives, with $x= \frac{\Delta v}{nV_T}$:
\begin{equation}
I_{\mathrm{out}}
=\frac{I_{\mathrm{in}}}{\big(1+e^{-x}\big)\big(1+e^{x}\big)}
=\frac{I_{\mathrm{in}}}{4}\,\operatorname{sech}^{2}\!\Big(\frac{x}{2}\Big),
\end{equation}
where $n$ is the subthreshold slope factor, $V_T$ is the thermal voltage and $\Delta v = V_1 - V_2$. 
% The current from a single differential pair is \begin{equation}
% I_1=I_{in}\frac{1}{1+\exp{\frac{-\Delta v_d}{nV_T}}}
% \end{equation}
% Cascading two differential achieves:
% \begin{equation}
% I_{out}=I_{in}\frac{1}{(1+\exp{\frac{-\Delta v_d}{nV_T}}){(1+\exp{\frac{\Delta v_d}{nV_T}}})}
% \end{equation}
% which is equal to 
% \begin{equation}
%   I_{\text{out}}(\Delta v)
%   \;=\;
%   \frac{I_{\text{in}}}{4}\,
% \operatorname{sech}^{2}\!\Big(\frac{\Delta v}{2\,n V_T}\Big)
% \end{equation}
Following a Taylor expansion about $\Delta v=0$, we have
$\operatorname{sech}^2 u = 1 - u^2 + O(u^4)$ with $u=\tfrac{\Delta v}{2nV_T}$,
while $\exp(-\gamma\,\Delta v^2)=1-\gamma\,\Delta v^2+O(\Delta v^4)$.
Matching the quadratic terms gives $\gamma=\tfrac{1}{4n^2V_T^2}$, so near the origin,
the cascaded pair provides a Gaussian approximation to the kernel:
% % Near the origin, $\operatorname{sech}^2 u = 1 - u^2 + O(u^4)$, while
% % $\exp(-u^2)= 1 -  u^2 + O(u^4)$.
% Matching the quadratic term gives the Gaussian approximation
\begin{equation}
I_{\mathrm{out}}(\Delta v)
\;\approx\;
\frac{I_{\mathrm{in}}}{4}\,\exp\!\Big(-\gamma\,\Delta v^2\Big),
\label{eq:circuit-gauss}
\end{equation}
% \blue{add the Taylor series part for more info on the approximation}
% \begin{equation}
%   I_{\text{out}}(\Delta v)
%   \;=\;
%   \frac{I_{\text{in}}}{4}\,
% \operatorname{sech}^{2}\!\Big(\frac{\Delta v}{2\,n V_T}\Big)
%   \;\approx\;
%   \frac{I_{\text{in}}}{4}\,
%   \exp\!\big[-\gamma\,\Delta v^{2}\big]
%   \label{eq:circuit-gauss}
% \end{equation}
% In this small-signal Gaussian approximation, 
where $\sigma$ denotes the standard deviation. 
Equivalently, we use $\gamma = 1/{2\sigma^{2}}\approx 1/(8n^2V_T^2)$, which aligns with $\gamma$ from \eqref{eq:rbf}.
% which is the standard RBF $\gamma$ parameter, 

The output current is sensed by an n-type transistor $Q_6$ whose gate and source are tied to $Q_4$.
In subthreshold, the drain current depends exponentially on $V_{GS}$ and only weakly on $V_{DS}$, so $Q_5$ produces a faithful, scaled replica of $Q_4$’s current.
Because we operate in subthreshold—where $I_D$ is only weakly dependent on $V_{DS}$—we use device ratioing for readout. Therefore, we undersize $Q_4$ (small $W/L$) to minimize loading, and set $Q_6$ with a $W/L$ similar to $Q_3$ so that the readout branch provides a larger copy of the kernel current for the next stage.

% \blue{need to mention the gamma scaling part by scaling the input and fixing the bias either here or in the experimental setup}
% \todo[inline]{still need to mention:
% 1. approximation through Taylor series for the first two components only, and how that affects the approximation --> this is only valid till vbias=300mv and afterwards the operating region is saturation instead of subthrehold for most of the mosfets.
% 2. Ibias affects the shape of the bell-shaped response
% 3. how to map different features (through equation 4 and pass the output current from one kernel to the input tail of the new kernel)
 
% }

\subsubsection{Kernel Product Across Dimensions}
\label{subsec:analog_kernel_product}
To handle $D$ dimensions, we implement a separable kernel by cascading $D$ identical one-dimensional cells per support vector. 
For support vector $j$ and dimension $d$, we form $\Delta v_{j,d}=V_{x(d)}-V_{s_{j,d}}$ and apply \eqref{eq:circuit-gauss} to that stage.
Because the output of stage $(d-1)$ drives the input of stage $d$ ($I_{\text{in},j}^{(d)}=I_{\text{out},j}^{(d-1)}$), the exponential factors multiply across stages. 
After $D$ stages, for each $j=1,\dots,m$ we obtain:
\begin{equation}
    I_{\text{out},j}^{(D)}
=\prod_{d=1}^D K_{j,d}=\frac{I_{\text{in},j}}{4^{D}}\,
\exp\!\Big(-\gamma \sum_{d=1}^{D} \Delta v_{j,d}^{2}\Big).
\end{equation}
When a large number of features is employed, the summation chain grows, resulting in current degradation that can no longer be reliably distinguished. 
Because the FE process is n-type transistors-only (no p-type devices), accurate current mirroring from n-type outputs is challenging. 
% and amplifier-based solutions are too power- and area-expensive. 
Therefore, we support up to five inputs in our design. 
% Therefore, we restrict the number of features by applying high-level feature selection, constraining the inputs to at most five. 
% Feature selection is widely used in ML applications to improve accuracy by automatically retaining only the most informative features.
% Moreover, as discussed in Section~\ref{sec:background:svms_alg}, RBF coefficients differ from linear ones in that the number of support vectors is unbounded and determined by training. 
% Thus, by applying feature selection, we not only reduce the input dimensionality but also decrease the overall size of the RBF classifier.

\subsubsection{Alpha Multiplier}
This block scales the kernel current by the dual coefficient through a controllable factor \(\alpha\in(0,1)\).
As shown in Fig~\ref{fig:analog_rbf_kernel}, it is implemented with a subthreshold differential pair \((Q_1,Q_2)\) with
diode-connected loads \((Q_3,Q_4)\). With control differential
\(\Delta V_{\alpha}=V_{\alpha+}-V_{\alpha-}\), the branch current follows a
logistic function:
\[
I_{j}^{(\alpha)} \;=\; I_{\text{out},j}^{(D)}\,
\frac{1}{1+\exp\!\big(\tfrac{\Delta V_{\alpha}}{nV_T}\big)}
\;\equiv\; I_{\text{out},j}^{(D)} \alpha(\Delta V_{\alpha})\, .
\]

\subsubsection{Signed Accumulation over Support Vectors}
As mentioned in (\ref{eq:svm_decision}), each support vector $j$ produces a nonnegative current $I_j=\alpha_j\,K_j(\mathbf{x})$.
% A \red{current–steering} switch, 
A switch controlled by the label $y_j\in\{+1,-1\}$, routes $I_j$ to the $+$ rail when $y_j=+1$ or to the $-$ rail when $y_j=-1$.
The two rails passively sum their incoming currents; 
% a differential readout (e.g., TIA or matched loads) computes the margin
% \begin{equation}
%     f(\mathbf{x})=\sum_{j=1}^{M} y_j\,\alpha_j\,K_j(\mathbf{x}),
% \end{equation}
Thereby realizing negative weights by accumulation on the negative rail.
% \blue{mentioning ideal comparator for comparing two currents}

Finally, the classifier’s score is passed through a comparator to obtain its sign, provides a digital output and seamlessly  feeds the digital decision making circuit without requiring any additional digitization, i.e., without an ADC.

% \TODO{@Maha: Please emphasize how the output of the analog RBF is binary, and why an ADC is not needed.}

% Further, it should be noted that the exponential function is implemented using a current-based design, similar to~\red{\cite{Kang2010gaussian}}. 

% \TODO{feature selction it is not something we only apply because it is a necessity of the analog architecture but also an area/power measurements against unconstrained everything parameters}
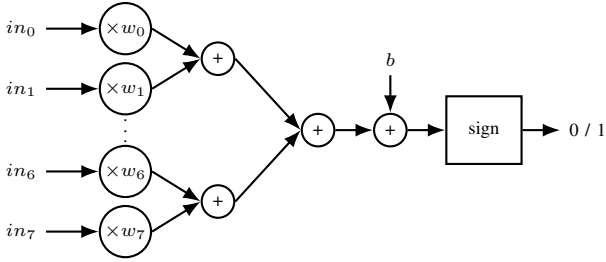
\begin{figure}[t!]
\centering
\begin{tikzpicture}[
  font=\scriptsize,
  node distance=1.2cm and 1.0cm,
  box/.style={draw, thick, minimum width=1cm, minimum height=.9cm, align=center},
  add/.style={draw, circle, thick, inner sep=2.5pt},
  mull/.style={draw, circle, thick, inner sep=1pt},
  arrow/.style={-Latex, thick}
]
\usetikzlibrary{arrows.meta,calc}

% Inputs
\node at (0,1.2) (x0) {$in_0$};
\node at (0,0.4) (x1) {$in_1$};
\node at (0,-0.7) (x6) {$in_6$};
\node at (0,-1.5) (x7) {$in_7$};

% Multipliers
\node[mull, right=0.7cm of x0] (mul0) {$\times w_0$};
\node[mull, right=0.7cm of x1] (mul1) {$\times w_1$};
\node[mull, right=0.7cm of x6] (mul6) {$\times w_6$};
\node[mull, right=0.7cm of x7] (mul7) {$\times w_7$};

% Connect inputs to multipliers
\draw[arrow] (x0.east) -- (mul0.west);
\draw[arrow] (x1.east) -- (mul1.west);
\draw[arrow] (x6.east) -- (mul6.west);
\draw[arrow] (x7.east) -- (mul7.west);

% Vertical dots between mul1 and mul6
\node at ($(mul1.south)!0.25!(mul6.north)$) (vdots) {$\vdots$};

% Adder tree
% \node[add, right=0.8cm of mul0] (add1) {+};
\node[add, right=1.0cm of $(mul0)!0.5!(mul1)$] (add1) {+}; % centered between mul0 and mul1
\node[add, right=1.0cm of $(mul6)!0.5!(mul7)$] (add2) {+};
% \node[add, right=1.0cm of add2] (add3) {+};
\node[add, right=1.1cm of $(add2)!0.5!(add1)$] (add3) {+}; % centered between mul0 and mul1

% \draw[arrow] (mul0.east) -- ++(0.5,0) |- (add1.west);
\draw[arrow] (mul1.east) -- (add1.west);
\draw[arrow] (mul7.east) -- (add2.west);
\draw[arrow] (mul6.east) -- (add2.west);

% Connections multipliers to adders
\draw[arrow] (mul0.east) -- (add1.west);
% \draw[arrow] (mul1.east) |- (add1.south);
% \draw[arrow] (mul7.east) -- ++(0.5,0) |- (add2.west);

% Connect adders
% \draw[arrow] (add1.east) -- ++(0.5,0) |- (add3.west);
% \draw[arrow] (add2.east) -- ++(0.5,0) |- (add3.west);
\draw[arrow] (add1.east) -- (add3.west);
\draw[arrow] (add2.east) -- (add3.west);

% \draw[arrow] (add2.east) -- (add3.west);

% Bias add
\node[add, right=0.5cm of add3] (biasadd) {+};
\node[above=0.5cm of biasadd] (b0) {$b$};
\draw[arrow] (b0.south) -- (biasadd.north);
\draw[arrow] (add3.east) -- (biasadd.west);

% Decision block
\node[box, right=.5cm of biasadd] (decision) {sign};
\draw[arrow] (biasadd.east) -- (decision.west);

% Output
\node[right=.5cm of decision] (out) {0 / 1};
\draw[arrow] (decision.east) -- (out.west);

\end{tikzpicture}
% \caption{Datapath of one linear SVM classifier with 8-inputs: inputs are multiplied by weights, accumulated with a bias, and the sign decision determines the binary output.}
\caption{Datapath of an 8-input linear SVM: weighted inputs are summed with bias, and the sign determines the output.}
\label{fig:digital}
\end{figure}
% \todo[inline]{dots}
\subsection{Digital Linear Classifier \& Decision-Making Logic}
\label{sec:digital}
In the digital domain, we implement the linear binary classifiers together with the multiclass decision-making logic of our mixed-signal SVM architecture.
Similar to state-of-the-art flexible SVM designs~\cite{Mubarik:MICRO:2020:printedml}, our linear kernels are designed as bespoke fully-parallel architectures, essentially implementing fixed-point MAC operations, as shown in \cref{fig:digital}.
Trained weights are quantized and multiplied with the quantized features as obtained from the ADC.
All products are computed in parallel and accumulated through an adder tree, followed by bias addition and sign evaluation. 
This yields a memory-less architecture, eliminating the considerable overhead of sequential elements.
% In FE, where nearly $99\%$ of the consumption is static power~\cite{Ozer:2019:Bespoke}, reducing sequential logic directly translates to significant energy savings~\red{\cite{afentaki:ISLPED25:flex_class}}.
Note, non-volatile memories remain scarce and costly in FE~\cite{ozer:nature2024:bendableRiscV}.
% for instance, the bendable RISC-V microprocessor in~\cite{ozer:nature2024:bendableRiscV} relied on off-chip memory.
In addition, weight and bias coefficients are hardwired within each linear classifier, following the bespoke design paradigm~\cite{ozer:nature2024:bendableRiscV}, substantially improving the area and power of flexible multipliers and adders compared to conventional (i.e., non-customized) implementations.

%% Bespoke paradigm
% We adopt a bespoke design paradigm in which the digital implementation is tailored to the target task and dataset (e.g., per-feature scaling, fixed-point formats, and per-classifier resource provisioning). 
% This level of specialization is enabled by the low non-recurring engineering (NRE) costs and rapid turn–around of FE, which make per-application customizations economically viable. 
% In contrast, such fine-grained tailoring is generally impractical in mainstream VLSI due to high mask costs and lengthy fabrication cycles. 
% As a result, FE allow us to co-design the model representation and the digital datapath to minimize memory, gate count, and switching activity while preserving accuracy.

%% Decision Making
The outputs of both digital linear and analog RBF classifiers are aggregated by the digital decision-making circuit to perform multiclass classification. 
Instead of majority voting as in \cite{Mubarik:MICRO:2020:printedml, Armeniakos:TCAD2023:cross}, decision-making is realized with an encoder, as shown in \figurename~\ref{fig:mixed_kernel_SVM}. 
Each binary classifier provides a one-bit output, and the encoder maps the collection of these binary outcomes directly to the corresponding class label. 
This hardware-efficient scheme removes the need for dedicated counter and \texttt{argmax} circuitry, thereby enabling the seamless integration of both digital and analog SVM components with minimal area and power overhead.
% low-latency inference and streamlined digital integration.
% The decision-making circuit aggregates the outcomes of all binary classifiers, and the final class label is assigned to the class with the majority of votes. 
% For $K$ classes, $\binom{K}{2}$ binary classifiers produce pairwise decisions. 
% The digital decision logic aggregates the outcomes by majority voting; in case of ties, margins are compared to select the final label.
% Integrating this logic in the digital domain keeps aggregation robust to noise and process variations.

% Overall, the digital domain provides an efficient and stable backbone for classification by combining linear kernel computation with multiclass decision logic, complementing the expressiveness offered by the analog RBF kernels.

\section{High-Level Mixed-Kernel Exploration}

\subsection{Analog RBF Modeling}
\label{sec:analog_model}
For seamless integration with digital simulations—without resorting to full mixed-signal co-simulation, which is complex and requires excessive circuit runs- we develop a high-level behavioral model of the analog RBF core. 
The model captures the Gaussian kernel and the $\alpha$ multiplier via parameters identified from targeted DC SPICE sweeps, enabling system-level accuracy evaluation. 
% To accurately assess the accuracy of the analog RBF classifier without repeatedly executing time–consuming circuit-level simulations for every classifier, we employ a high-level behavioral model.
This approach increases automation at a high level while preserving final accuracy through calibration against measured curves.
Below, the modeling of each component in the analog RBF classifier (see \cref{subsec:analog} and \cref{fig:analog_rbf_kernel}) is described.

% , as parameter extraction and evaluation are fully scripted. 
% The model parameters are identified from SPICE characterizations and then reused across datasets.

\subsubsection{Gaussian kernel Modeling}
We model the hardware kernel by running a DC analysis, sweeping the differential input
$\Delta v$, and using the resulting samples $I_{\mathrm{out}}(\Delta v)$. However, in order
to measure the $\gamma$ of the simulated design, we fit an \emph{ideal Gaussian} to the
measured transfer characteristic—specifically estimating $A_0$, $\gamma_0$, and $\mu$:
\begin{equation}
    I_{\mathrm{out}}(\Delta v)\;\approx\;A_0\,\exp\!\big[-\gamma_0\,(\Delta v-\mu)^2\big].
\end{equation}
We then use the SPICE data together with the fitted $\gamma_0$ in the behavioral modeling.

We fix the operating point at a specific bias \(V_b\). To achieve different RBF widths \(\gamma^*\),
we apply input scaling relative to the measured \(\gamma_0\) and the separable kernel is then evaluated as:
\begin{equation}
s_\gamma=\sqrt{\gamma^\ast/\gamma_0}, \qquad K(\Delta v)=\prod_{d=1}^{D} K(s_\gamma\,\Delta v).
\end{equation}

\subsubsection{Kernel product across dimensions}
We multiply the per-dimension kernels to realize a separable $D$-dimensional kernel. No additional approximation is introduced here: the current is passed from one stage to the next; only the input scaling $s_\gamma$ differs.

\subsubsection{Alpha multiplier}
We model the dual-coefficient magnitude $\alpha\in(0,1)$ with a logistic curve, fitted to SPICE by a DC sweep of the control differential $\Delta V_\alpha$ versus the measured ratio $\alpha=I_{\mathrm{out}}/I_{\mathrm{in}}$. The fit yields $(x_0,s)$, and in software we map a desired $\alpha$ to its control value via:
\begin{equation}
    \Delta V_\alpha = x_0 + s\,\ln\!\Big(\frac{1}{\alpha}-1\Big)
\end{equation}

\subsection{Mixed-Kernel Mapping \& Training}
\label{sec:highlevel}
\begin{algorithm}[t]
\caption{Mixed-kernel SVM selection process.}
\label{alg:highlevel-ovo}
\begin{algorithmic}[1]
\small
\Require Labeled dataset $\mathcal{D}$ with classes $\{c_1,\dots,c_K\}$
\State $\mathcal{P}\!\leftarrow\!\{(c_i,c_j)\mid i<j\}$ \Comment{All OvO pairs}
\State $\mathcal{K}\!\leftarrow\!\emptyset$ \Comment{Chosen kernel per pair}
\State $\mathcal{C}\!\leftarrow\!\emptyset$ \Comment{Trained classifier per pair}

\ForAll{$(c_i,c_j)\in\mathcal{P}$}
  \State $\mathcal{D}_{ij}\!\leftarrow\!\{(x,y)\in\mathcal{D}\mid y\in\{c_i,c_j\}\}$
  \State Train linear $C^{\text{lin}}_{ij}$ and RBF $C^{\text{rbf}}_{ij}$ on $\mathcal{D}_{ij}$
  \State $A_{\text{lin}}\!\leftarrow\!\mathrm{acc}(C^{\text{lin}}_{ij}),\ \ A_{\text{rbf}}\!\leftarrow\!\mathrm{acc}(C^{\text{rbf}}_{ij})$
  \If{$A_{\text{rbf}} > A_{\text{lin}}$}
     \State $k_{ij}\!\leftarrow\!\text{RBF}$,\ \ $C_{ij}\!\leftarrow\!C^{\text{rbf}}_{ij}$
  \Else
     \State $k_{ij}\!\leftarrow\!\text{Linear}$,\ \ $C_{ij}\!\leftarrow\!C^{\text{lin}}_{ij}$
  \EndIf
  \State $\mathcal{K}\!\leftarrow\!\mathcal{K}\cup\{k_{ij}\}$,\ \ $\mathcal{C}\!\leftarrow\!\mathcal{C}\cup\{C_{ij}\}$
\EndFor
\State \textbf{Decision step (encoder):} For input $x$, compute binary outputs $b_{ij}(x)\!\in\!\{0,1\}$ from all $C_{ij}\!\in\!\mathcal{C}$; obtain $\hat{y}$
\State \textbf{Evaluation:} $\mathrm{Acc}(\text{SVM}_{\text{mixed}})\!\leftarrow\!\mathrm{acc}(\hat{y};\mathcal{D})$
\State \textbf{Outputs:} $\mathcal{K}$ (kernel map), $\mathcal{C}$ (mixed-kernel classifiers)
\end{algorithmic}
\end{algorithm}

% \todo[inline]{maybe mention that for the digital different regularizations where also testing that corresponds to more zero weights and therefore removing multipliers(?) since analog implementation is an approximation of the actual rbf classifier it would make sense}
% \TODO{@Florentia: I think some reasoning for hardware cost needs to be mentioned. How this selection improves the overall cost for example. We only emphasize accuracy}

% We propose a separation-based strategy for optimally mapping each SVM binary classifier to its optimal kernel (i.e., linear or RBF), thus determining its implementation (i.e., digital or analog).
% \todo[inline]{Explain where analog model is integrated in the exploration}
% \vspace{-0.7mm}

Leveraging that SVM OvO multi-class classification decomposes the final classification to distinct binary classification problems, we propose a separation strategy to assign each binary classifier to the optimal kernel type, RBF or linear.
Our objective is to maximize the accuracy of each binary classifier---enabling superior model-level accuracy compared to single-kernel state-of-the-art approaches---whilst minimizing the number of RBF kernels.
% , and thus the overall SVM model, which allows our technique to achieve superior accuracy compared to single-kernel state-of-the-art approaches (as later shown in \cref{sec:eval}).

Algorithm~\ref{alg:highlevel-ovo} summarizes our mapping strategy.
For a $K$-class task, we enumerate all $\binom{K}{2}$ pairs of classes $(c_i,c_j)$ and extract the corresponding binary subset for each pair.
Each pair is first trained with a linear classifier, and its accuracy is used to gauge classification difficulty. 
If the accuracy is sufficient, the pair remains linear; if not, it is reassigned to an RBF kernel.
In this way, most classifiers remain lightweight and digital, while only the challenging ones are realized as analog RBFs. 
This selective allocation preserves the accuracy benefits of RBFs where they matter most, while minimizing their overhead and maintaining an overall efficient design.
% For every pair, we train the two candidate binary SVM classifiers, a linear- and an RBF-based one, keeping the one with higher accuracy in separating its two assigned classes.
% We then keep the kernel that performs better in terms of accuracy, integrating it into the mixed-kernel model.
% This greedy approach deconstructs SVM classification into pairwise sub-tasks, handling each individual classifier as a standalone model, without loss of generality.
% Prioritizing accuracy enables us to fully harvest the computational benefits of RBF kernels, whilst inserting digital logic wherever linear classifiers are competitive.
% Prioritizing accuracy with low power and area, RBF kernels are employed only when they deliver higher accuracy than linear, minimizing their area and power overhead while keeping the overall design lightweight.
% The objective is to preserve the accuracy benefits of RBF where they are superior, while limiting the number of analog blocks—costly and variability-sensitive in FE—by assigning linear models wherever they are competitive.
Finally, after all classifiers are extracted, they are integrated within a unified SVM model, alongside the decision-making logic for outputting the predicted class, per the OvO scheme.
Floating-point RBF classifiers are replaced with our hardware-accurate high-level analog model of \cref{sec:analog_model}.
% multi-class prediction is executed via the standard OvO algorithm using majority voting among all classifiers. 
The exploration therefore outputs: 
(i) a pairwise kernel assignment map distinguishing linear (digital) from RBF (analog) 
% classifiers---directly informing the mixed-signal architectural mapping---
and (ii) the mixed-kernel SVM model, ready for system-level inference.
% Once the exploration is completed, the highest-accuracy model is selected and evaluated using the analog implementation.
\section{Results \& Analysis}
\label{sec:evaluation_results}

\subsection{Experimental Setup}
\label{sec:experimental_setup}

\subsubsection{Software Setup}
We evaluate our mixed-kernel SVM approach over 3 datasets from the UCI ML repository~\cite{Dua:2019:uci}, as they are well-suited for sensor-based FE applications~\cite{Mubarik:MICRO:2020:printedml}.
% , allowing direct comparisons with the state of the art.
Datasets include Balance Scale (Balance/Bal.), Seeds and Vertebral 3 Columns (Vertebral/V3C). 
Sensor data are normalized within [0,1], and any non-sensor (categorical) features are removed during pre-processing. 
% Each dataset is randomly split into training and testing sets using a $70\%/30\%$ ratio. 
We split each dataset into 70\% training and 30\% testing.
Feature selection is applied before training to align the input dimensions with our analog hardware constraints (see \cref{subsec:analog_kernel_product}), limiting up to 5 input features.
% SVM training and feature selection is performed using Scikit-learn.
Scikit-learn is used for SVM training and feature selection.

\begin{table}[t]
\centering
\caption{Component dimensions of analog RBF classifier}
\label{tab:RBF_sizing}
\setlength{\tabcolsep}{6pt}
\begin{tabular}{l l}
\toprule
\multicolumn{2}{c}{\textbf{Gaussian Kernel }}\\
\midrule
$Q_1$--$Q_3$, $Q_6$ & W = 40\,\si{\micro\meter},\ L = 0.6\,\si{\micro\meter} \\
$Q_4$                         & W = 1\,\si{\micro\meter},\ L = 0.6\,\si{\micro\meter} \\
$Q_5$                         & W = 20\,\si{\micro\meter},\ L = 1.2\,\si{\micro\meter} \\
% $Q_6$                         & W = 40\,\si{\micro\meter},\ L = 0.6\,\si{\micro\meter} \\
$R_1$= \SI{10}{\mega\ohm}
 & W = 0.6\,\si{\micro\meter},\ L = 28.5\,\si{\micro\meter} \\
$R_2$ = \SI{4.28}{\mega\ohm}
 & W = 0.6\,\si{\micro\meter},\ L = 12.2\,\si{\micro\meter} \\
\hline
\addlinespace[2pt]
\multicolumn{2}{c}{\textbf{Alpha Multiplier}}\\
\midrule
$Q_1$--$Q_4$             & W = 40\,\si{\micro\meter},\ L = 0.6\,\si{\micro\meter} \\

\bottomrule
\end{tabular}
\end{table}

\subsubsection{Hardware Setup}
Sensory inputs are uniformly quantized to 4-bit fixed-point precision.
\red{For linear SVMs, weights and biases are quantized following~\cite{Armeniakos:TCAD2023:cross} to preserve accuracy.
For digital RBF SVMs support vectors and dual coefficients are quantized to ensure sufficient precision.}
For analog simulations and digital circuit synthesis and simulation, Cadence Spectre simulator, Synopsys Design Compiler T-2021.06 and VCS T-2022.03 are used, respectively. 
Our designs are mapped to PragmatIC's Gen3 FlexIC 1.0.0 PDK at \SI{1.5}{\volt}~\cite{flexic_gen3}.
% For mapping our designs to the flexible technology Gen3 PragmatIC FlexIC 1.0.0 PDK is used at \SI{1.5}{\volt} and \SI{25}{\celsius}~\cite{flexic_gen3}.
% For analog simulations, we use the Cadence Spectre simulator with the Gen3 PragmatIC FlexIC 1.0.0 PDK~\cite{flexic_gen3}.
% \red{The supply voltage for the analog is set to \SI{1.5}{\volt}},
The supply voltage for the analog is set to \SI{1}{\volt}, regulated down from the digital supply voltage.
The dimensions of all components in our analog RBF classifier design are reported in \cref{tab:RBF_sizing}, \red{for the comparator, we estimated power and area from the comparator designed in~\cite{shatta2025featuretoclassifiercodesign}}.
Both analog and digital target performance at \SI{2}{\hertz}.
It should be noted that the target applications of FE are inherently low-throughput, and therefore the chosen operating frequency aligns with the typical speed requirements of such systems~\cite{ozer:nature2024:bendableRiscV}.

% \begin{figure}[t]
%   \centering
%   % Left panel: TABLE (fits to half-column)
%   \begin{minipage}[t]{0.49\columnwidth}
%   % \label{metrics_validation}
%     \vspace{0pt} % top align
%     \centering
%     \scriptsize
%     \setlength{\tabcolsep}{3pt}

%     \resizebox{\linewidth}{!}{%
%     \begin{tabular}{@{}>{\raggedright\arraybackslash}p{.58\linewidth}
%                     S[table-format=1.4]
%                     S[table-format=1.3]@{}}
%       \toprule
%       \textbf{Component} & \textbf{nRMSE} & \textbf{$r$} \\
%       \midrule
%       Gaussian kernel (at $V_b=0.30$ V) & 0.0218 & 0.997 \\
%       Product across dims ($D=3$)       & 0.0117 & 0.998 \\
%       Alpha multiplier (logistic fit)    & 0.0003 & 0.999 \\
%       \bottomrule
%     \end{tabular}}
%     % \vspace{11pt}
%     { (a) SPICE vs.\ ideal reference}
%   \end{minipage}
%   % Right panel: FIGURE
%   \begin{minipage}[t]{0.49\columnwidth}
%     \vspace{0pt}
%     \centering
%     \includegraphics[width=\linewidth]{figures/kernel_fit.pdf}
%     % \vspace{1pt}
%     {\footnotesize (b) Gaussian kernel fit vs.\ SPICE.}
%   \end{minipage}

%   \caption{Analog RBF Validation (a) SPICE vs.\ ideal metrics and (b) Gaussian kernel comparison.}
%   \label{fig:table_plus_kernel}
% \end{figure}

\begin{figure}[t]
  \centering
  % Left panel: TABLE
  \begin{subfigure}[t]{0.49\columnwidth}
    \vspace{0pt} % top align
    \centering
    \scriptsize
    \setlength{\tabcolsep}{3pt}
    \resizebox{\linewidth}{!}{%
    \begin{tabular}{@{}>{\raggedright\arraybackslash}p{.58\linewidth}
                    S[table-format=1.4]
                    S[table-format=1.3]@{}}
      \toprule
      \textbf{Component} & \textbf{nRMSE} & \textbf{$r$} \\
      \midrule
      Gaussian kernel (at $V_b=0.30$ V) & 0.0218 & 0.997 \\
      Product across dims ($D=3$)       & 0.0117 & 0.998 \\
      Alpha multiplier (logistic fit)   & 0.0003 & 0.999 \\
      \bottomrule
    \end{tabular}}
    \subcaption{SPICE vs.\ ideal reference}
  \end{subfigure}
  \hfill
  % Right panel: FIGURE
  \begin{subfigure}[t]{0.49\columnwidth}
    \vspace{0pt} % force top alignment
    \centering
    \includegraphics[width=\linewidth]{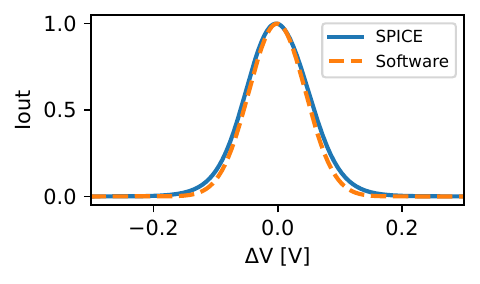}
    \vspace{-4.3ex}
    \subcaption{Gaussian kernel fit vs.\ SPICE}
  \end{subfigure}

  \caption{Analog RBF validation: (a) SPICE vs.\ ideal metrics and (b) Gaussian kernel comparison.}
  \label{fig:table_plus_kernel}
\end{figure}

\subsection{Analog RBF validation}
First, we validate the behavior of our analog RBF classifier design, and its comprising components.
Specifically, we compare the obtained SPICE simulation traces against an \emph{ideal} software reference for the following components:
the Gaussian kernel cell, the product across dimensions, and the alpha-multiplier. 
For each case, we report the normalized root mean square error (nRMSE) and the correlation between SPICE and the ideal outputs.
The results are summarized in Fig.~\ref{fig:table_plus_kernel}(a).
Overall, we observe excellent agreement with the theoretical reference, with all blocks exhibiting very low nRMSE and near-unity correlation ($r\!\approx\!1$).
Fig.~\ref{fig:table_plus_kernel}(b) also presents the I-V curve of the Gaussian kernel in comparison to the ideal software function, providing a visual confirmation of this alignment.
Thus, our analog implementation provides accurate (and well-correlated) outputs, faithful to the ideal RBF implementation outlined in \cref{subsec:analog}.

\subsection{System Evaluation}
Next, we evaluate our mixed-signal and mixed-kernel SVMs in terms of accuracy and hardware efficiency. 
Specifically, we focus our analysis on the following:
(i) comparison of single–kernel baselines (digital linear, digital RBF, and our mixed-signal mixed-kernel SVM),
(ii) analysis of analog vs.\ digital RBF efficiency, 
% and (iii) a breakdown of the mixed design into linear and RBF contributions to quantify area and power impact.
(iii) a breakdown of linear and RBF contributions in the mixed design to quantify area and power.

\begin{table}[!t]
\setlength{\tabcolsep}{1.5pt}  % default is 6pt
\centering
\caption{Evaluation against single-kernel state-of-the-art.}
\label{tab:acc_area_power_comp}
\begin{threeparttable}
\begin{tabular}{
    l l
    % S[table-format=3.0]   % Software Acc. (%)
    S[table-format=2.0]   % Circuit Acc. (%)
    S[table-format=2.4]   % Total Area (mm^2)
    S[table-format=1.3]   % Total Power (mW)
    c
}
\toprule
\textbf{Dataset} &
\textbf{Design} &
% {\thead{Software\\Acc. (\%)}} &
{\thead{\textbf{Circuit} \\\textbf{Acc.} (\%)}} &
{\thead{\textbf{Total Area}\\(\si{\milli\meter^2})}} &
{\thead{\textbf{Total Power}\\(\si{\milli\watt})}} &
{\thead{\textbf{RBF/linear}\\\textbf{classifier ratio}}}\\

\midrule
% \multicolumn{4}{c}{\textbf{Balance}} \\ \midrule
\multirow{3}{*}{Bal.}
& Linear (digital) & 92 &  0.024 & 0.004 & 0/3\\
& RBF (digital)    & 93 &  13.400 & 2.230 & 3/0\\
& \textbf{Ours (mixed)}  & 92 & 0.062 & 0.081 & 1/2 \\
\midrule
% \multicolumn{4}{c}{\textbf{Seeds}} \\ \midrule
\multirow{3}{*}{Seeds}
& Linear (digital) & 92 & \multicolumn{1}{c}{0.067} & \multicolumn{1}{c}{0.011} & 0/3\\
& RBF (digital)    & 95 & 7.000  & 1.190  & 3/0\\
& \textbf{Ours (mixed)}  & 95 & 0.125 & 0.092  &1/2\\
\midrule
% \multicolumn{4}{c}{\textbf{Vertebral}} \\ \midrule
\multirow{3}{*}{V3C}
& Linear (digital) & 69 & \multicolumn{1}{c}{0.092} & \multicolumn{1}{c}{0.014} & 0/3\\
& RBF (digital)    & 83 & 5.600  & 0.960  & 3/0\\
& \textbf{Ours (mixed)}  & 89 & 0.108 & 0.088  & 2/1\\
% \midrule
% \multicolumn{4}{c}{\textbf{\red{Wesad E4}}} \\ \midrule
% Linear (digital) & 69  & \multicolumn{1}{c}{0.092} & \multicolumn{1}{c}{0.014} \\
% RBF (digital)    & 83 & 5.600  & 0.960  \\
% \textbf{Ours (mixed)}  & 89                     & 0.108 & 0.088  \\
\bottomrule
\end{tabular}
% \begin{tablenotes}[flushleft]
% \scriptsize “--” indicates not applicable or not measured (software-only).
% \end{tablenotes}
\end{threeparttable}
\end{table}
% \todo[inline]{circuit acc only}

% \subsubsection{Comparison with Single–Kernel Baselines}
We compare our SVMs against the state-of-the-art SVM designs, which focus solely on single–kernel linear digital implementations.
In addition, we design against single-kernel digital RBF SVMs, aiming for more comprehensive comparisons.
% \Red{even though they are not adopted by the state of the art due to their excessive hardware overhead.}
Table~\ref{tab:acc_area_power_comp} reports the results in terms of circuit accuracy, total area, and power.
Compared to purely-linear SVMs, our mixed-kernel designs offer superior classification accuracy, with an average increase of 7.7\%, reaching up to 20\% for Vertebral.
This comes at the cost of moderate overhead, since digital linear classifiers yield lower area and power cost compared to analog RBF ones---by 2.5x and 12.4x, respectively.
In contrast, compared to single-kernel digital RBF SVMs, our designs offer orders-of-magnitude higher efficiency, with an average area and power gain of \red{108x} and \red{$17\times$}, respectively.
At the same time, they achieve comparable accuracy, ranging from a minor 1\% drop for Balance, to equal values for Seeds, up to a 6\% gain for Vertebral, showcasing the advantages of mixed-kernel SVMs.
% For Seeds, the linear baseline delivers $92\%$ with the lowest cost, whereas the digital RBF attains $95\%$. 
% The mixed–kernel matches the $95\%$ accuracy while reducing area and power by $56\times$ and $16\times$, respectively, compared to the digital RBF.  
% For Vertebral, the linear model is the most efficient but reaches only $69\%$. 
% The digital RBF improves accuracy by $14\%$, while our mixed–kernel achieves a further increase of $6\%$ over single-kernel RBF, with $52\times$ less area and $11\times$ less power than the digital RBF.  
% For Balance, the digital linear SVM achieves $92\%$ accuracy with the smallest footprint.
% A digital RBF improves accuracy by only one percentage point, but at a prohibitive cost.
% Our mixed–kernel design maintains $92\%$ accuracy while requiring $216\times$ less area and $32\times$ less power than the digital RBF.  
% It should be mentioned that Balance is the smallest dataset with only 3 classifier, both the single-kernel linear while the linear kernels in our mixed-kernel approach resulted with zero coefficients the area thus as we will show later is its area is dominated by the very much more complexer rbf. 
% Overall, digital linear SVMs prove the most area/power–efficient at the cost of some accuracy loss;
% digital RBF designs are the most accurate, but come with prohibitively high hardware overheads.
\emph{Overall, our mixed–kernel and mixed-signal SVMs offer the best accuracy–efficiency trade–off, surpassing state-of-the-art accuracy of linear SVMs while providing huge hardware gains over digital RBF designs.}

% \begin{figure}[!t]
% \centering
% \includegraphics[width=\columnwidth]{figures/mixed_kernel_area_power_subplot.pdf}
% \caption{Area and power breakdown between analog/digital in our mixed-signal SVMs of \cref{tab:acc_area_power_comp}.}
% \label{fig:breakdown}
% \end{figure}

\begin{figure}[t]
  \centering
  \begin{tikzpicture}
  \begin{groupplot}[
    group style={group size=2 by 1, horizontal sep=5em},
    width=0.45\columnwidth,
    height=0.37\columnwidth,
    ybar stacked,
    ymin=0, ymax=1,
    x=18pt,
    ytick={0,0.25,0.5,0.75,1},
    yticklabel style={font=\footnotesize},
    symbolic x coords={Bal.,Seeds,V3C},
    xtick=data,
    xtick style={draw=none},
    xticklabel style={rotate=0, anchor=center, font=\footnotesize, },
    tick align=outside,
    title style={font=\footnotesize},
    legend style={at={(1.3,1.08)}, anchor=south, legend columns=-1, draw=none, font=\footnotesize},
    legend cell align=center
  ]

  % ---- Plot 1: Normalized AREA ----
  \nextgroupplot[
    ylabel={Area (norm.)},      % <- custom y-label
    enlarge x limits=0.5
  ]
  \addplot coordinates {(Bal.,0.09) (Seeds,0.89) (V3C,0.62)};
  \addplot coordinates {(Bal.,0.91) (Seeds,0.11) (V3C,0.38)};
  \title{sd}
  \legend{Digital, Analog};

  % ---- Plot 2: Normalized POWER ----
  \nextgroupplot[
    ylabel={Power (norm.)},     % <- custom y-label
    enlarge x limits=0.5
  ]
  \addplot coordinates {(Bal.,0.01) (Seeds,0.21) (V3C,0.1111)};
  \addplot coordinates {(Bal.,0.99) (Seeds,0.79) (V3C,0.8889)};

  \end{groupplot}
  \end{tikzpicture}
\vspace{-2ex}
\caption{Normalized area and power breakdown between analog/digital in our mixed-signal SVMs of \cref{tab:acc_area_power_comp}.}
\label{fig:breakdown}
\vspace{-2ex}
\end{figure}
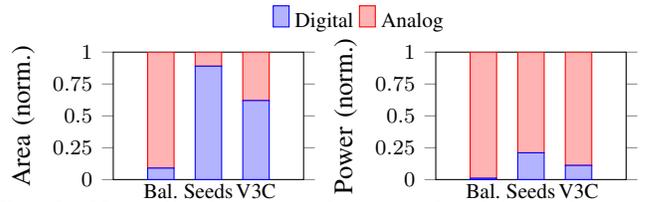

Note, we do not provide a direct comparison against single-kernel RBF-based SVMs implemented purely in the analog domain, since we are the first to design analog RBF classifiers in FlexIC technology.
In addition, our mixed-signal designs deliver more favorable trade-offs given the hardware constraints of FE applications.
We achieve high accuracy---within 1\% of the software accuracy---while the analog RBF classifier is costlier than its digital linear counterpart.

% We use a digital RBF as the hardware-cost baseline, configured with the same support vectors, $\gamma$, and dual coefficients $(\alpha,y)$ as the analog flow for a fair comparison. 
% For \emph{Balance}, the digital RBF consumes 13.4 mm$^2$ and 2.23 mW versus 0.14 mm$^2$ and 0.095 mW for the analog—i.e., $95\times$ area and $23.5\times$ power. 
% For \emph{Seeds}, it is 7.0 mm$^2$/1.19 mW vs. 0.04 mm$^2$/0.078 mW—$175\times$ area and $15.3\times$ power. 
% For \emph{Vertebral}, 5.6 mm$^2$/0.960 mW vs. 0.097 mm$^2$/0.091 mW—$57.7\times$ area and $10.5\times$ power. 
% These gaps show that realizing the RBF kernel in analog is essential for hardware efficiency, which is, on average, \textbf{109x} \textbf{and} \textbf{16x} \textbf{area and power efficient}, respectively.
% \subsubsection{Comparison Analog vs. Digital RBF}
Next, we perform a direct comparison between digital RBF classifiers from (Table~\ref{tab:acc_area_power_comp}) against our analog designs.
Our analog classifiers are on average 109$\times$ more area-efficient and 16$\times$ more power-efficient, highlighting the importance of analog realization, and enabling for the first time RBF-based SVMs in FlexIC technology.
\label{sec:eval}
% \todo[inline]{The analog model has a larger accuracy loss compared to software than I expected in most cases. Can we justify this? For example, does the difference increase with the percentage of RBF classifiers?}

% \todo[inline]{- make sure to show area, power, accuracy results for all digital linear, all digital RBF, and your mix solution. }

% \subsubsection{Mixed–Kernel Classifier}
% \include{figures/stucked_ratios}
Finally, we present an area and power breakdown of our mixed-kernel solutions of \cref{tab:acc_area_power_comp} between their analog and digital components.
Fig.~\ref{fig:breakdown} presents the normalized results per the studied datasets.
Overall, the digital end (i.e., linear classifiers and decision making) accounts for 54\% of the area, mainly due to the separation nature of our exploration, which aims to minimize the count of RBF blocks.
Interestingly, the analog RBF dominates the total area for Balance by nearly 90\%. 
This occurs because the digital linear component converged to zero or power of 2 weights, which in this bespoke architecture directly translates to hardware savings both area and power (99\% static power in FE~\cite{ozer:nature2024:bendableRiscV} by removing the corresponding multipliers.
According to \cref{fig:breakdown}(b), the analog RBF classifiers dominate the total power consumption by 89\%, on average.
This reflects the inherently higher power overhead of analog computation in FE, despite its area efficiency.

\section{Conclusion}
\label{sec:conclusion}
% Flexible electronics (FE) are emerging as a pathway to low-cost, conformal computing for sensor-driven applications.
% Yet, implementing ML algorithms in FE remains costly due to limited circuit efficiency.
% We propose a mixed-kernel SVM combining digital linear kernels with analog RBF.
% Our evaluations show up to $6\%$ higher accuracy than linear baselines, while reducing area and power by $108\times$ and $17\times$ versus fully digital RBFs.
% Our results highlight that our mixed-kernel mixed-signal methodology, offer a practical and scalable route to hardware-efficient intelligence in FE systems.
Flexible electronics (FE) are emerging as a pathway to low-cost, conformal computing for sensor-driven applications. 
However, implementing ML algorithms in FE remains costly due to the limited efficiency of complex circuits.
In this work, we propose a mixed-kernel SVM that combines digital linear kernels with analog RBF.
Our evaluations highlight that our mixed approach achieves, on average $7.7\%$ higher accuracy than linear baselines with minimal hardware overhead, while reducing area and power by up to $108\times$ and $17\times$, respectively, compared to fully digital RBFs. 
These results demonstrate that analog RBFs, when coupled with a lightweight digital pipeline, provide a practical and scalable route to hardware-efficient intelligence in FE systems.

\section*{Acknowledgements}

\small
This work is partially supported by the European Research Council (ERC) (Grant No. 101052764),  and  co-funded by the H.F.R.I call “Basic research Financing (Horizontal support of all Sciences)” under the National Recovery and Resilience Plan “Greece 2.0” (H.F.R.I. Project Number: 17048).

\clearpage
\balance
\bibliographystyle{IEEEtran}
\bibliography{refs/7_refsAbrv, refs/7_references, refs/references_iscas}

@inproceedings{tahoori:ETS2025:PFE,
  title={Computing with Printed and Flexible Electronics},
  author={Tahoori, Mehdi B and Ozer, Emre and Zervakis, Georgios and Balaskas, Konstantinos and Pal, Priyanjana},
  booktitle={2025 IEEE European Test Symposium (ETS)},
  pages={1--9},
  year={2025},
  organization={IEEE}
}

@article{burges1998tutorial,
  title   = {A Tutorial on Support Vector Machines for Pattern Recognition},
  author  = {Burges, Christopher J. C.},
  journal = {Data Mining and Knowledge Discovery},
  volume  = {2},
  number  = {2},
  pages   = {121--167},
  year    = {1998},
  publisher = {Springer},
  doi     = {10.1023/A:1009715923555}
}

@article{Burges:DMKD:1998,
  author    = {Christopher J. C. Burges},
  title     = {A Tutorial on Support Vector Machines for Pattern Recognition},
  journal   = {Data Mining and Knowledge Discovery},
  volume    = {2},
  number    = {2},
  pages     = {121--167},
  year      = {1998},
  doi       = {10.1023/A:1009715923555}
}

@article{Smola:StatsComp:2004,
  author    = {Alex J. Smola and Bernhard Sch{\"o}lkopf},
  title     = {A Tutorial on Support Vector Regression},
  journal   = {Statistics and Computing},
  volume    = {14},
  number    = {3},
  pages     = {199--222},
  year      = {2004},
  doi       = {10.1023/B:STCO.0000035301.49549.88}
}

@article{Cavallaro:TCAS:2010,
  author    = {Joseph R. Cavallaro and Hyuk-Jae Lee and Kyounghoon Yang},
  title     = {Analog Circuit Implementations of Kernel-Based SVM Classifiers},
  journal   = {IEEE Transactions on Circuits and Systems I: Regular Papers},
  volume    = {57},
  number    = {6},
  pages     = {1441--1454},
  year      = {2010},
  doi       = {10.1109/TCSI.2009.2036828}
}

@inproceedings{Li:ISCAS:2017,
  author    = {Xiaolong Li and H. Jiang and R. Shafik and et al.},
  title     = {Low-Power Analog Implementation of Radial Basis Function Kernel for SVM},
  booktitle = {IEEE International Symposium on Circuits and Systems (ISCAS)},
  year      = {2017},
  pages     = {1--4},
  doi       = {10.1109/ISCAS.2017.8050842}
}

@article{xu:nature2014:epidermal,
  author    = {Sheng Xu and Yonggang Huang and John A. Rogers},
  title     = {Epidermal Electronics},
  journal   = {Science},
  year      = {2014},
  volume    = {344},
  number    = {6179},
  pages     = {70--72},
  doi       = {10.1126/science.1239381}
}

@misc{pragmatic:whitepaper:sustainability,
  author       = {{PragmatIC Semiconductor}},
  title        = {Sustainability Through Flexible Electronics},
  howpublished = {\url{https://www.pragmaticsemi.com/sustainability/}},
  year         = {2023},
  note         = {Whitepaper}
}

@article{afifi2020fpga,
  title={FPGA implementations of SVM classifiers: A review},
  author={Afifi, Shereen and GholamHosseini, Hamid and Sinha, Roopak},
  journal={SN computer science},
  volume={1},
  number={3},
  pages={133},
  year={2020},
  publisher={Springer}
}

@INPROCEEDINGS{afifi:IECBES2016:fgpa_SVM_melanoma,
  author={Afifi, Shereen and GholamHosseini, Hamid and Sinha, Roopak},
  booktitle={2016 IEEE EMBS Conference on Biomedical Engineering and Sciences (IECBES)}, 
  title={A low-cost FPGA-based SVM classifier for melanoma detection}, 
  year={2016},
  volume={},
  number={},
  pages={631-636},
  doi={10.1109/IECBES.2016.7843526}
}

@article{alimisis2023hybridsvm,
  author    = {Vassilis Alimisis and Georgios Gennis and Marios Gourdouparis and Christos Dimas and Paul P. Sotiriadis},
  title     = {A Low-Power Analog Integrated Implementation of the Support Vector Machine Algorithm with On-Chip Learning Tested on a Bearing Fault Application},
  journal   = {Sensors},
  volume    = {23},
  number    = {8},
  pages     = {3978},
  doi       = {10.3390/s23083978},
  year      = {2023}
}

@article{tcsi2009analogsvm,
  author    = {Kyunghee Kang and Tadashi Shibata},
  title     = {Analog Gaussian-Kernel Support Vector Machine: A Fully Trainable VLSI Approach},
  journal   = {IEEE Transactions on Circuits and Systems I: Regular Papers},
  volume    = {57},
  number    = {7},
  pages     = {1513--1524},
  doi       = {10.1109/TCSI.2009.2024904},
  year      = {2009}
}

@INPROCEEDINGS{sertaridis:ISCAS2025:SequentialSVM_1_mac,
  author={Sertaridis, Ilias and Besias, Spyridon and Afentaki, Florentia and Balaskas, Konstantinos and Zervakis, Georgios},
  booktitle={IEEE International Symposium on Circuits and Systems (ISCAS)}, 
  title={Compact Yet Highly Accurate Printed Classifiers Using Sequential Support Vector Machine Circuits}, 
  year={2025},
  volume={},
  number={},
  pages={}
}

@article{besias:DATE2025:Sequential_N_mac,
  title={Late Breaking Results: Energy-Efficient Printed Machine Learning Classifiers with Sequential SVMs},
  author={Besias, Spyridon and Sertaridis, Ilias and Afentaki, Florentia and Balaskas, Konstantinos and Zervakis, Georgios},
  booktitle={Design Automation and Test in Europe (DATE)}, 
  year={2025}
}

@article{ozer:nature2024:bendableRiscV,
  title={Bendable non-silicon RISC-V microprocessor},
  author={Ozer, Emre and Kufel, Jedrzej and Prakash, Shvetank and Raisiardali, Alireza and Kindgren, Olof and Wong, Ronald and Ng, Nelson and Jausseran, Damien and Alkhalil, Feras and Kong, David and others},
  journal={Nature},
  pages={1--6},
  year={2024},
  publisher={Nature Publishing Group UK London}
}

@misc{Dua:2019:uci ,
author = "Dua, Dheeru and Graff, Casey",
year = "2017",
title = "{UCI} Machine Learning Repository",
institution = "University of California, Irvine, School of Information and Computer Sciences" }

@ARTICLE{Kang2010gaussian,
  author={Kang, Kyunghee and Shibata, Tadashi},
  journal={IEEE Transactions on Circuits and Systems I: Regular Papers}, 
  title={An On-Chip-Trainable Gaussian-Kernel Analog Support Vector Machine}, 
  year={2010},
  volume={57},
  number={7},
  pages={1513-1524},
  doi={10.1109/TCSI.2009.2034234}}

@misc{flexic_gen3,
  author = {Pragmatic},
  title = {FlexIC {P}latform {G}en3},
  year = {2025},
  howpublished = {\url{https://www.pragmaticsemi.com/foundry/flexic-platform-gen-3}}
}

@misc{shatta2025featuretoclassifiercodesign,
      title={Invited Paper: Feature-to-Classifier Co-Design for Mixed-Signal Smart Flexible Wearables for Healthcare at the Extreme Edge}, 
      author={Maha Shatta and Konstantinos Balaskas and Paula Carolina Lozano Duarte and Georgios Panagopoulos and Mehdi B. Tahoori and Georgios Zervakis},
      year={2025},
      eprint={2508.19637},
      archivePrefix={arXiv},
      primaryClass={eess.SP},
      url={https://arxiv.org/abs/2508.19637}, 
}

@IEEEtranBSTCTL{IEEEexample:BSTcontrol,
    CTLuse_forced_etal       = {yes},
    CTLmax_names_forced_etal = {5},
    CTLnames_show_etal       = {1},
    CTLdash_repeated_names = "no"
}

@inproceedings{Afentaki:DATE2024:embedding,
  title={Embedding Hardware Approximations in Discrete Genetic-based Training for Printed MLPs},
  author={Afentaki, Florentia and Hefenbrock, Michael and Zervakis, Georgios and Tahoori, Mehdi B},
  booktitle={2024 Design, Automation \& Test in Europe Conference \& Exhibition (DATE)},
  pages={1--6},
  year={2024},
  organization={IEEE}
}

@inproceedings{Afentaki:ICCAD23:hollistic,
  author={Afentaki, Florentia and Saglam, Gurol and Kokkinis, Argyris and Siozios, Kostas and Zervakis, Georgios and Tahoori, Mehdi B.},
  booktitle={2023 IEEE/ACM International Conference on Computer Aided Design (ICCAD)}, 
  title={Bespoke Approximation of Multiplication-Accumulation and Activation Targeting Printed Multilayer Perceptrons}, 
  year={2023},
  volume={},
  number={},
  pages={1-9},
  doi={10.1109/ICCAD57390.2023.10323613}
}

@inproceedings{Mubarik:MICRO:2020:printedml,
  author={Mubarik, M. H. and Weller, D. D. and Bleier, N. and Tomei, M. and Aghassi-Hagmann, J. and Tahoori, M. B. and Kumar, R.},
  booktitle={Annu. Int. Symp. Microarchitecture (MICRO)}, 
  title={Printed Machine Learning Classifiers}, 
  year={2020},
  volume={},
  number={},
  pages={73-87},
  doi={10.1109/MICRO50266.2020.00019}
}

@inproceedings{Armeniakos:DATE2022:axml,
  author={Armeniakos, Giorgos and Zervakis, Georgios and Soudris, Dimitrios and Tahoori, Mehdi B. and Henkel, Jörg},
  booktitle={Design Automation and Test in Europe (DATE)},
  title={Cross-Layer Approximation For Printed Machine Learning Circuits}, 
  year={2022},
  volume={},
  number={},
  pages={190-195},
  doi={10.23919/DATE54114.2022.9774689}
}

@article{Armeniakos:TCAD2023:cross,
  author={Armeniakos, Giorgos and Zervakis, Georgios and Soudris, Dimitrios and Tahoori, Mehdi B. and Henkel, Jörg},
  journal={IEEE Transactions on Computer-Aided Design of Integrated Circuits and Systems},
  title={Model-to-Circuit Cross-Approximation For Printed Machine Learning Classifiers}, 
  year={2023},
  volume={},
  number={},
  pages={1-1},
  doi={10.1109/TCAD.2023.3258668}
}

@article{Armeniakos:TC2023:codesign,
  author={Armeniakos, Giorgos and Zervakis, Georgios and Soudris, Dimitrios and Tahoori, Mehdi B. and Henkel, Jörg},
  journal={IEEE Transactions on Computers}, 
  title={Co-Design of Approximate Multilayer Perceptron for Ultra-Resource Constrained Printed Circuits}, 
  year={2023},
  volume={},
  number={},
  pages={1-8},
  doi={10.1109/TC.2023.3251863}
}

@article{smartpackaging2022,
title = {Application of inkjet-printing technology in developing indicators/sensors for intelligent packaging systems},
journal = {Current Opinion in Food Science},
volume = {46},
pages = {100868},
year = {2022},
issn = {2214-7993},
doi = {https://doi.org/10.1016/j.cofs.2022.100868},
url = {https://www.sciencedirect.com/science/article/pii/S2214799322000704},
author = {Xiaoyu Luo}
}

@article{disposable:JSNB:2023,
title = {Screen-printed graphene-carbon ink based disposable humidity sensor with wireless communication},
journal = {Sensors and Actuators B: Chemical},
volume = {374},
pages = {132731},
year = {2023},
issn = {0925-4005},
doi = {https://doi.org/10.1016/j.snb.2022.132731},
url = {https://www.sciencedirect.com/science/article/pii/S0925400522013740},
author = {Ajay Beniwal and Priyanka Ganguly and Akshaya Kumar Aliyana and Gaurav Khandelwal and Ravinder Dahiya}
}

@Article{healthcare:Nanoscale:2024,
author ="Zhou, Kemeng and Ding, Ruochen and Ma, Xiaohao and Lin, Yuanjing",
title  ="Printable and flexible integrated sensing systems for wireless healthcare",
journal  ="Nanoscale",
year  ="2024",
volume  ="16",
issue  ="15",
pages  ="7264-7286",
publisher  ="The Royal Society of Chemistry",
doi  ="10.1039/D3NR06099C",
url  ="http://dx.doi.org/10.1039/D3NR06099C",
}

@article{Wearable:acssensors:2019,
author = {Agarwala, Shweta and Goh, Guo Liang and Dinh Le, Truong-Son and An, Jianing and Peh, Zhen Kai and Yeong, Wai Yee and Kim, Young-Jin},
title = {Wearable Bandage-Based Strain Sensor for Home Healthcare: Combining 3D Aerosol Jet Printing and Laser Sintering},
journal = {ACS Sensors},
volume = {4},
number = {1},
pages = {218-226},
year = {2019},
doi = {10.1021/acssensors.8b01293},
}

@article{wearable:adma:2022,
author = {Gao, Jiuwei and Fan, Yubo and Zhang, Qingtian and Luo, Lei and Hu, Xiaoqi and Li, Yue and Song, Juncai and Jiang, Hanjun and Gao, Xiaoyu and Zheng, Lu and Zhao, Wu and Wang, Zhenhua and Ai, Wei and Wei, Yuan and Lu, Qianbo and Xu, Manzhang and Wang, Yongtian and Song, Weitao and Wang, Xuewen and Huang, Wei},
title = {Ultra-Robust and Extensible Fibrous Mechanical Sensors for Wearable Smart Healthcare},
journal = {Advanced Materials},
volume = {34},
number = {20},
pages = {2107511},
doi = {https://doi.org/10.1002/adma.202107511},
url = {https://onlinelibrary.wiley.com/doi/abs/10.1002/adma.202107511},
year = {2022}
}

@article{pressuresensor:research:2022,
author = {Yue Li  and Yuan Wei  and Yabao Yang  and Lu Zheng  and Lei Luo  and Jiuwei Gao  and Hanjun Jiang  and Juncai Song  and Manzhang Xu  and Xuewen Wang  and Wei Huang },
title = {The Soft-Strain Effect Enabled High-Performance Flexible Pressure Sensor and Its Application in Monitoring Pulse Waves},
journal = {Research},
number = {},
year = {2022},
doi = {10.34133/research.0002},
}

@article{bodytemperature:sna:2020,
title = {High sensitivity flexible paper temperature sensor and body-attachable patch for thermometers},
journal = {Sensors and Actuators A: Physical},
year = {2020},
issn = {0924-4247},
doi = {https://doi.org/10.1016/j.sna.2020.112205},
url = {https://www.sciencedirect.com/science/article/pii/S0924424720305495},
author = {Jin-Woo Lee and Younguk Choi and Jaehoon Jang and Se-Hyuk Yeom and Wanghoon Lee and Byeong-Kwon Ju},
}

@INPROCEEDINGS{Henkel:ICCAD2022:expedition,
  author={Henkel, Jörg and Li, Hai and Raghunathan, Anand and Tahoori, Mehdi B. and Venkataramani, Swagath and Yang, Xiaoxuan and Zervakis, Georgios},
  booktitle={International Conference On Computer Aided Design (ICCAD)}, 
  title={Approximate Computing and the Efficient Machine Learning Expedition}, 
  year={2022},
  volume={},
  number={},
  pages={1-9},
  doi={}}

@INPROCEEDINGS{Mrazek:ICCAD2024,
author = {Mrazek, Vojtech and Kokkinis, Argyris and Papanikolaou, Panagiotis and Vasicek, Zdenek and Siozios, Kostas and Tzimpragos, Georgios and Tahoori, Mehdi and Zervakis, Georgios},
booktitle={International Conference On Computer Aided Design (ICCAD)}, 
year = {2024},
pages = {},
title = {Evolutionary Approximation of Ternary Neurons for On-sensor Printed Neural Networks}
}

\end{document}